\documentclass[final,5p,authoryear,times,twocolumn]{elsarticle}
\usepackage{natbib}
\usepackage{color}
\usepackage[latin1]{inputenc}
\usepackage{graphics}
\usepackage{amssymb}
\usepackage{overpic}
\journal{Icarus}
\begin{document}
\begin{frontmatter}

\title{Outgassing of icy bodies in the solar system - I. The sublimation of hexagonal water ice through dust layers} 
\author[igep]{B. Gundlach}
\ead{b.gundlach@tu-bs.de}
\author[igep]{Yu. V. Skorov}
\author[igep]{J. Blum}
\address[igep]{Institut für Geophysik und extraterrestrische Physik, Technische Universität Braunschweig, \\Mendelssohnstr. 3, D-38106 Braunschweig, Germany}
\begin{abstract}
Our knowledge about the physical processes determining the activity of comets were mainly influenced by several extremely successful space missions (Giotto, Deep Space I, Stardust, Deep Impact and EPOXI), the predictions of theoretical models and the results of laboratory experiments. However, novel computer models should not be treated in isolation but should be based on experimental results and should be verified and calibrated by experimental work. Therefore, a new experimental setup was constructed to investigate the temperature dependent sublimation properties of hexagonal water ice and the gas diffusion through a dry dust layer covering the ice surface. We show that this experimental setup is capable to reproduce known gas production rates of pure hexagonal water ice. The reduction of the gas production rate due to an additional dust layer on top of the ice surface was measured and compared to the results of another experimental setup in which the gas diffusion through dust layers at room temperature was investigated. We found that the relative permeability of the dust layer is inversely proportional to its thickness, which is also predicted by theoretical models. However, the measured absolute weakening of the gas flow was smaller than predicted by models. This lack of correspondence between model and experiment may be caused by an ill-determination of the boundary condition in the theoretical models, which further demonstrates the necessity of laboratory investigations. Furthermore, the impedance of the dust layer to the ice evaporation was found to be similar to the impedance at room temperature, which means that the temperature profile of the dust layer is not influencing the reduction of the gas production. Finally, we present the results of an extended investigation of the sublimation coefficient, which is an important factor for the description of the sublimation rate of water ice and, thus, an important value for thermophysical modeling of icy bodies in the solar system. The achieved results of this laboratory investigations demonstrate that experimental works are essential for the understanding of the origin of cometary activity.
\end{abstract}
\begin{keyword}
Comets, dust \sep Comets, dynamics \sep Comets, nucleus \sep Ices.
\end{keyword}
\end{frontmatter}

\section{Introduction}
Over the last decade, our understanding of the physical processes determining the activity of comets has deepened considerably. The main driving forces behind this progress were extremely successful space missions to comets: Deep Space I, Stardust, Deep Impact and, most recently, EPOXI. These missions delivered many new and sometimes unexpected information, which urgently requires the development of new theoretical models as well as new laboratory investigations. The construction of new computer models should not be treated in isolation but should be based on experimental results and should be verified and calibrated by experimental work.
\par
The most extended and challenging series of the comet simulation experiments (known as KOSI) were performed in the 1980s and 1990s using the German Space Agency's hardware facilities and the space simulation chamber in Cologne. The experiments dealt with the energy balance of insolated ice-dust mixture, the evolution of structure and composition of the sample, ice sublimation and emission of dust particles. A review of the main results and their critical analysis can be found, for example in \citet{Laemmerzahl1995} and \citet{Sears1999}. Although the KOSI experiments provided new insights into the morphology and physical behavior of comet analogs, the complicated setup of the experiments significantly impeded a quantitative analysis of the observations and did not cause a corresponding development of theoretical models.
\par
In order to avoid such complexity and to create a solid experimental foundation for further theoretical investigations, we decided to carry out a set of small-scale experiments performed under conditions approximating those in the environment of cometary nuclei. The basic idea of these experiments is to use accurate methods for the investigation of the physical behavior of well-defined cometary analog materials. In this paper, we present the results of laboratory investigations on the sublimation of a solid hexagonal water ice and the reduction of sublimation rate by a porous non-volatile dust layer covering the icy sample.
\par
Similar explorations have been executed in several laboratories since the 1990s. Speaking about such experiments, one should first mention the extended series performed in the laboratories of Tel-Aviv University and the University of Graz. The Tel-Aviv group focused mainly on the study of amorphous water ice, its crystallization and the release of other volatiles trapped by the amorphous water ice. In the recent publications \citep{BarNun2008, PatEl2009}, they presented fascinating results obtained for samples of gas-laden amorphous ice, which well reproduce the findings by the Deep Impact mission. The Graz group studied the energy and mass transport in both, solid and porous, crystalline water ice, including phenomena occurring when the icy sample was covered by a non-volatile layer. Their last experiments dealt with the exploration of the so-called solid-state greenhouse effect, which may play an important role in the energy balance of icy surfaces in the solar system \citep{Kaufmann2007}. All these achievements are impressive. However, it is surprising that until now we have not obtained  accurate systematic values of the effective sublimation rate of water ice and the weakening of the sublimation beneath a porous dust layer.
This paper is organized the following way: in Sect. 2, we describe the experimental approach for the determination of the sublimation rate of hexagonal water ice and the impedance to the evaporation exerted by overlaying dust layers; Sect. 3 summarizes our experimental results for the sublimation of hexagonal water ice, the gas diffusion through dust layers, and the temperature dependence of the sublimation coefficient; finally, Sect. 4 indicates possible applications to icy bodies in the solar system.

\section{Experimental} \label{Experimental}
To investigate the temperature dependent sublimation properties of hexagonal water ice and the gas diffusion inside a dry dust layer covering the ice surface, a new experiment consisting of an ice-dust sample under high vacuum condition and cryogenic temperature was constructed (sublimation experiment) in our institute. The gas diffusion through identical dust layers was also measured with another experimental setup (diffusion experiment) in order to verify the obtained experimental results. This second experiment was performed at room temperature to avoid condensation and evaporation of gas molecules inside the dust layer. In this section, the technical setups, the experimental procedures and the sample preparation are described.

\subsection{Sublimation experiment} \label{Sublimation experiment}
\begin{figure}[t]
\centering
\includegraphics[angle=0,width=1.00\columnwidth]{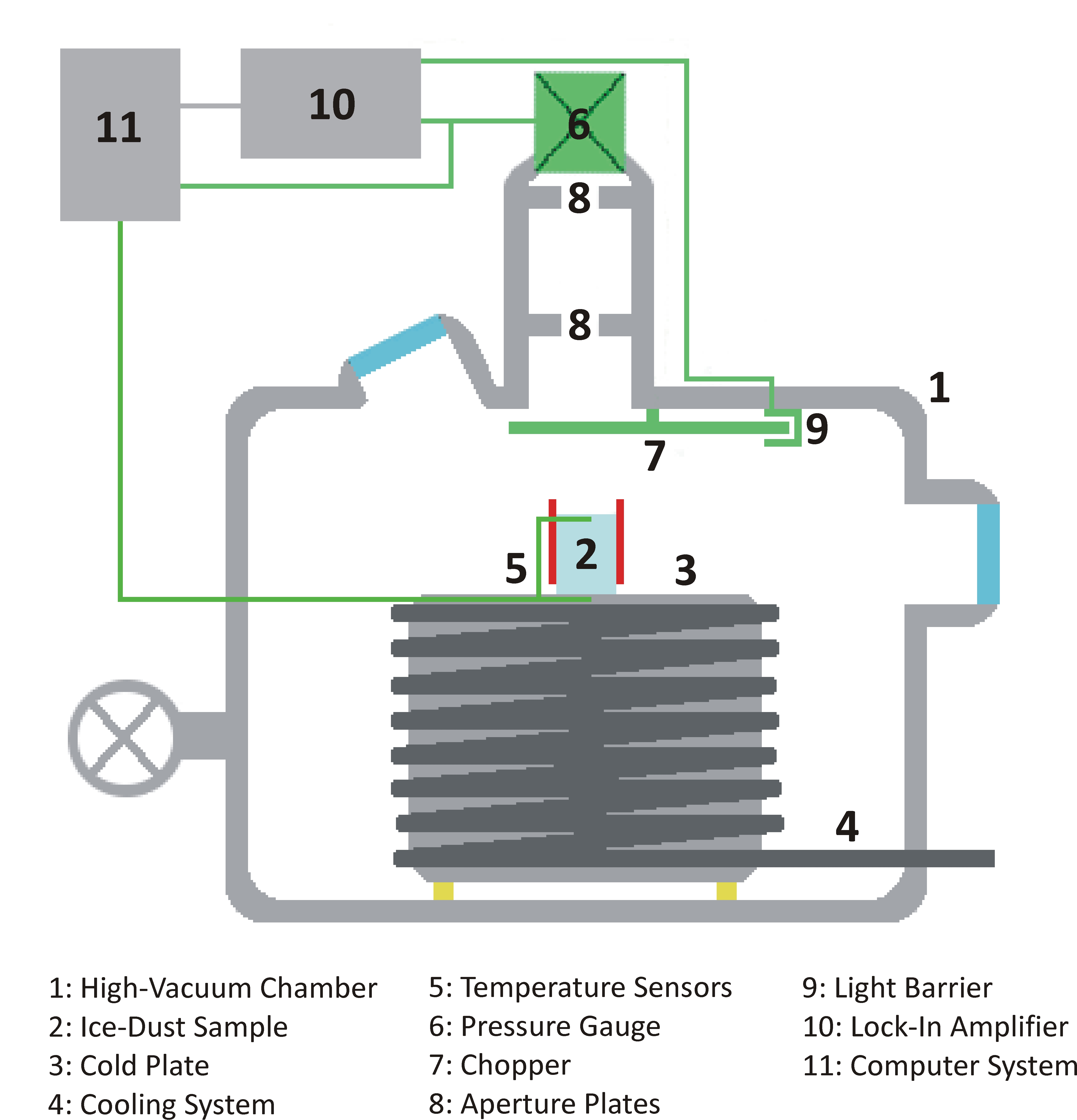}
\caption{Schematic diagram of the sublimation experiment. Sublimated water molecules from the sample's surface are detected using a pressure gauge together with a rotating chopper wheel.}
\label{experiment}
\end{figure}

The design of the sublimation experiment is shown in Fig. \ref{experiment}. The measurements with this setup were performed in a high-vacuum chamber (1) at pressures below $10^{-2}\,\mathrm{Pa}$. An ice-dust sample (2, see Sect. \ref{Sample preparation}) was frozen onto a cold plate (3), which was connected to the cooling system (4). The entire system was cooled down to approximately $\sim110\,\mathrm{K}$ with liquid nitrogen ($\mathrm{N_2}$). Temperature sensors (5) inside the ice-dust sample and the cold plate enabled temperature measurements of the ice surface and the cooling system. With this experimental setup, the gas production rate of the ice-dust sample can be measured, using a pressure gauge (6) positioned $54\,\mathrm{cm}$ above the sample's surface and a rotating chopper wheel (7), which was used to periodically interrupt the flux of gas molecules. Therefore, the rotating chopper wheel was located between the sample and the sensor. A complete description of the gas flux measurement is given in Sect. \ref{Measurement of the gas flux}. The sublimated water molecules were collimated into a narrow beam by two aperture plates (8) to eliminate the background gas flux. The signals of the pressure gauge and the light barrier (9), which was used to measure the rotational frequency of the chopper wheel, were input into a lock-in amplifier (10) to determine the sublimation rate of the ice-dust sample (see Sect. \ref{Measurement of the gas flux}). Measurements can be performed automatically every two seconds by a computer system (11). The results of the data analysis are discussed in Sect. \ref{Results}.
\par
In total, two different types of sublimation experiments were performed. For the calibration of the experimental setup, the gas production rate of hexagonal water ice was measured and compared with previous theoretical and experimental works (see Sect. \ref{Sublimation of water ice}). Afterwards, the ice-dust samples were used to investigate the gas diffusion inside dry dust layers covering the sample's surface and their influence on the sublimation properties of water ice (see Sect.\ref{Gas diffusion in the porous dust layer}).
\par
Before the start of a measurement sequence, the high-vacuum chamber was filled with nitrogen gas to avoid condensation of humidity onto the ice samples. Afterwards, liquid nitrogen was led through the pipes of the cooling system in order to cool down the cold plate. At a temperature of $\sim255\,\mathrm{K}$, a water droplet was used to freeze the sample onto the cold plate. Then, the high-vacuum chamber was evacuated from atmospheric pressure to a pressure below $10^{-2} \, \mathrm{Pa}$. To exclude intermolecular collisions between the gas molecules, the measurements of the gas production rate and the temperature were performed below $1.14 \times 10^{-2}\, \mathrm{Pa}$, when the mean free path of the gas molecules, $\lambda$, was larger than the distance between the sample's surface and the pressure gauge. A detailed explanation of the gas production rate measurements is given in Sect. \ref{Measurement of the gas flux}. The thickness of the dust layer was the only parameter which was varied among the sublimation experiments.

\subsection{Diffusion experiment}\label{Diffusion experiment}
\begin{figure}[ht]
\centering
\includegraphics[angle=0,width=0.70\columnwidth]{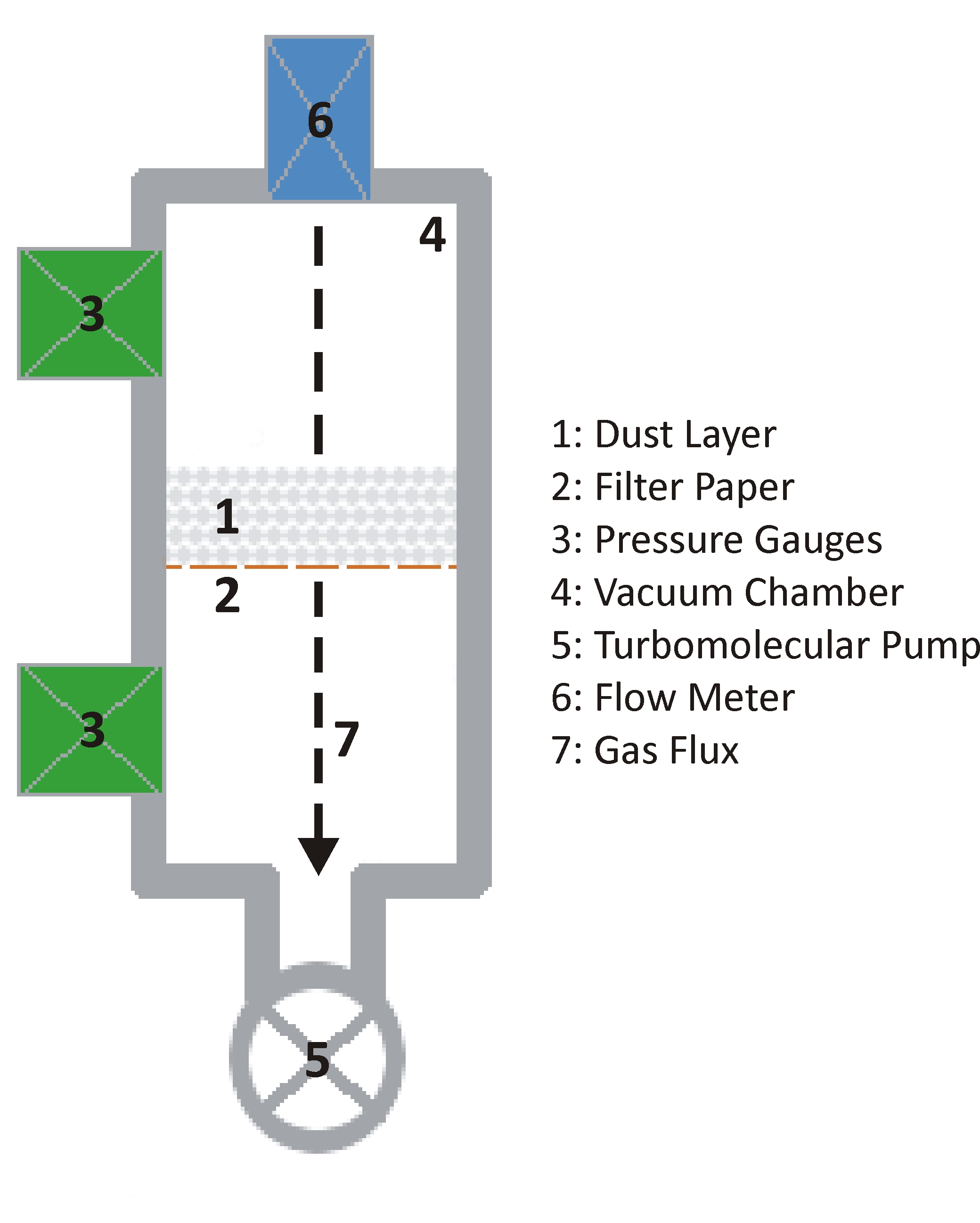}
\caption{Design of the diffusion experiment. The gas flux (dashed line) was forced to penetrate through the dust layer by the flow meter and the vacuum pump.}
\label{DiffusionExperimentPlot}
\end{figure}

To investigate the gas diffusion through dust layers (see Sect. \ref{Sample preparation}) with a different method, a second experiment was constructed (diffusion experiment, see Fig. \ref{DiffusionExperimentPlot}). A dust layer of given properties (1) was positioned onto a filter paper (2), between two pressure gauges (3) inside a vacuum chamber (4), which was evacuated to a pressure of $\sim 10^{-1} \, \mathrm{Pa}$ by a turbomolecular pump (5). A flow meter (6) enabled a precise control of the gas flux (7) through the dust layer. The dust layers investigated with this experiment were not cooled to prevent condensation and evaporation and to guarantee an investigation of the pure gas-permeability process.
\par
For each experiment, four different gas fluxes (air) through the dust layer $\phi_{df}$ were adjusted with the flow meter: $0.56 \times 10^{-3} \, \mathrm{Pa\, m^3 \, s^{-1}}$, $0.94 \times 10^{-3} \, \mathrm{Pa\, m^3 \, s^{-1}}$, $1.31 \times 10^{-3} \, \mathrm{Pa\, m^3 \, s^{-1}}$ and $1.69 \times 10^{-3} \, \mathrm{Pa\, m^3 \, s^{-1}}$. These settings resulted in mean pressures ranging from $8.4 \, \mathrm{Pa}$ to $105 \, \mathrm{Pa}$. After the gas flux adjustment, the pressure above and beneath the dust layer and the filter paper was measured and the permeability of the dust layer was computed (see Sect. \ref{Gas diffusion in the porous dust layer}). The permeability of the filter paper without dust layer was used for the normalization of the obtained results (see Sect. \ref{Gas diffusion in the porous dust layer}). The results of these measurements are then compared with the obtained data from the sublimation experiment in Sect. \ref{Gas diffusion in the porous dust layer}.

\subsection{Sample preparation} \label{Sample preparation}
\begin{figure}[t]
\centering
\begin{overpic}[angle=0,width=0.49\columnwidth]{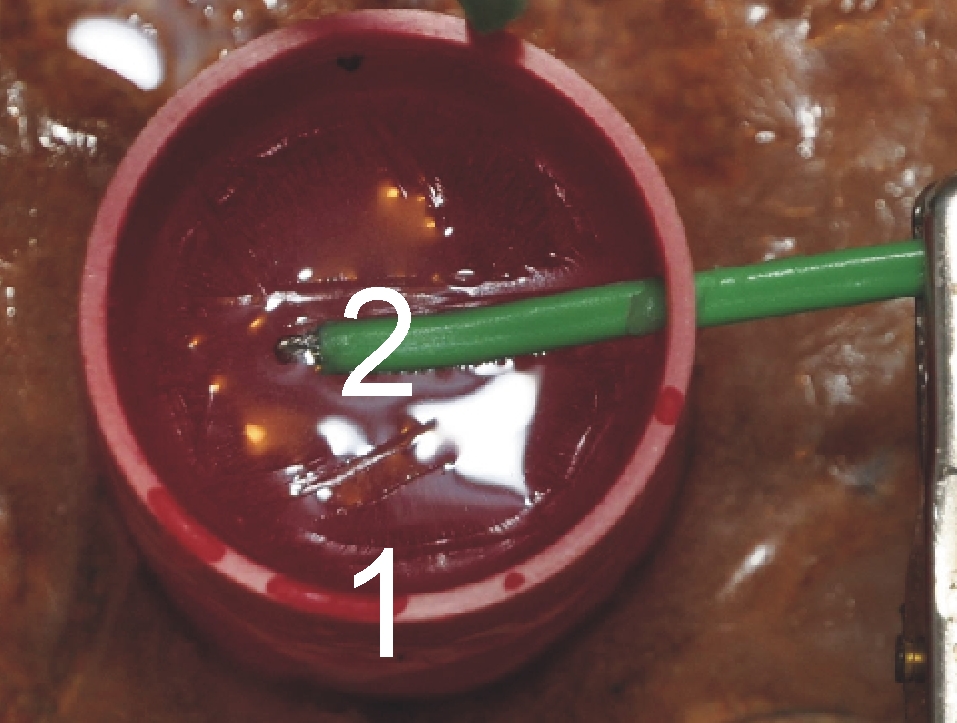}
\put(3,67){\color{white}a)\color{black}}
\end{overpic}
\begin{overpic}[angle=0,width=0.49\columnwidth]{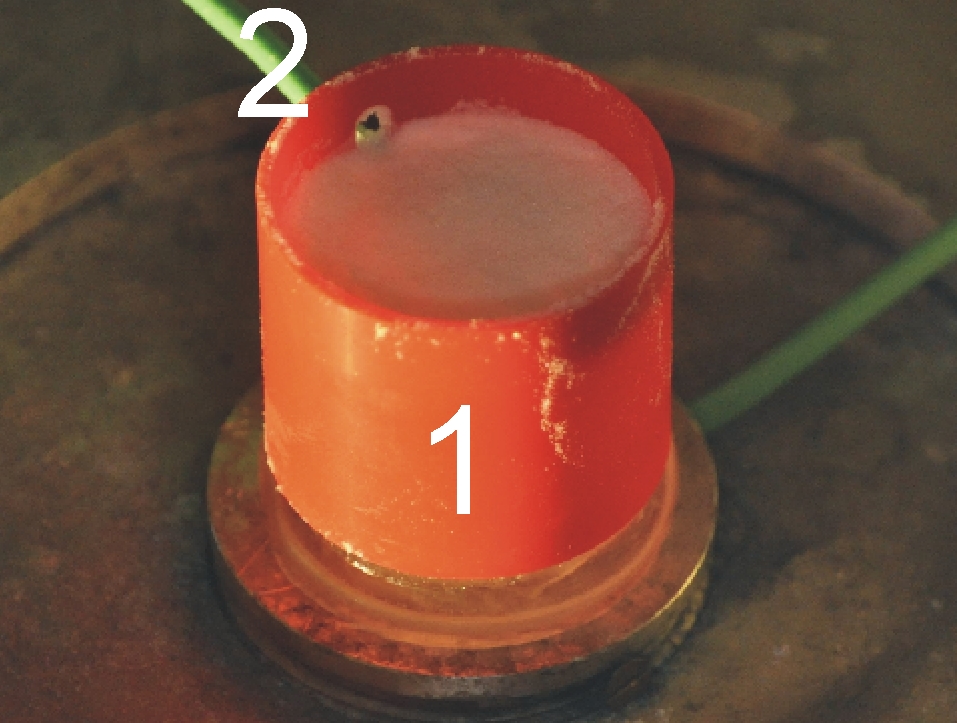}
\put(3,67){\color{white}b)\color{black}}
\end{overpic}
\caption{Pictures of two different cylindrical, hexagonal-water-ice samples inside the sample holder (1). For the temperature measurement, a thermocouple (2) was frozen into the ice close to the surface (a). Picture (b) shows a porous, dry dust layer composed of $(54.7\pm11.0)\,\mathrm{\mu m}$ sized silica spheres, which was added on top of the ice sample (ice-dust sample).}
\label{sample}
\end{figure}

In the sublimation experiments, cylindrical, hexagonal-water-ice samples with a diameter of $25\,\mathrm{mm}$ and a height of $30\,\mathrm{mm}$ were used. The ice samples (see Fig. \ref{sample}) were produced with distilled water that was frozen within a cylindrical sample holder at a temperature of $\sim255\,\mathrm{K}$. To minimize the heat flow through the sample holder, a material with a low heat conductivity (polyvinyl chloride, $\kappa = 0.15 \, \mathrm{W\,K^{-1}\,m^{-1}}$) and a small wall thickness ($1 \, \mathrm{mm}$) in relation to its diameter was used. Due to the geometry of the sample holder, sublimation from the samples's side walls was prevented. Both, the heat flow through the sample holder and the sublimation of molecules from the sample's side walls, can have a considerable effect on the energy balance of the sample and, thus, on the surface temperature of the ice. For the temperature measurement of the icy surface, a thermocouple was frozen into the ice sample at a distance of $<3$~mm from the surface.
\par
In eleven sublimation experiments (S1-S11, see Table \ref{TableResultsSublimationExperiment}) a dry dust layer composed of $(54.7\pm11.0)\,\mathrm{\mu m}$ sized silica spheres, was added on top of the ice sample (ice-dust samples). To estimate the height of the dust layer, the mass of the dust layer was measured before and after the experiments. The size distribution of these particles was measured with a microscope. Maximum compaction of the dust resulted in a constant porosity of $\psi = (40.8\pm0.3) \, \%$ for the entire set of experiments. The porosity was determined by measuring the mass of water, which was filled into the voids of the dust. Furthermore, the compacted silica dust layer was also used in 17 diffusion experiments (D1-D17, see Table \ref{TableResultsDiffusionExperiment}).
\par
To investigate the gas diffusion through a monolayer of dust, 133 glass spheres with a monodisperse diameter of $2\,\mathrm{mm}$, were positioned on the ice surface (S0) as well as on the filter paper (D0). This arrangement resulted in a porosity of $\psi = 43.4 \, \%$, which is very close to the porosity of the dust layer.

\subsection{Measurement of the gas production rate} \label{Measurement of the gas flux}
\begin{table*}[t!h]
\begin{center}
    \small
    \caption{Results of different empirical measurements of the saturation pressure of water ice. The parameters $a_1$ and $a_2$ refer to Eq. \ref{eq4}. }\vspace{1mm}
    \begin{tabular}{lr@{$\,$}lr@{$\pm$}l}
        \hline
        Reference                                                       & \multicolumn{2}{c}{$a_1$ [$\times10^{12}\,\mathrm{Pa}$]}                       & \multicolumn{2}{c}{$a_2$ [K]}           \\
        \hline
         \citet{Fanale1984}                                             & \multicolumn{2}{c}{$3.56$}                    & \multicolumn{2}{c}{$6141.667$}    \\
         \citet{Mauersberger2003}: $T > 169 \mathrm{K}$                 & $3.44$   & $^{+0.09}_{-0.08}$                       & $6132.9\,$ & $\,1.8$              \\
         \citet{Mauersberger2003}: $T < 169 \mathrm{K}$                 & $758.58$ & $^{\tiny{+312.9}}_{-221.5}$                    & $7044\,$   & $\,60$               \\
         This work                                                      & $3.23$   & $^{+0.73}_{-0.38}$                      & $6134.6\,$  & $\,17.0$        \\
        \hline
    \end{tabular}
     \label{SublimationPressureParameter}
     \end{center}
\end{table*}
The measurement of the gas production rate during the sublimation experiments was conducted by using a Bayard-Alpert-type pressure gauge positioned above the sample's surface and a rotating chopper wheel, which was located between the sample and the sensor. Two apertures with a diameter of $25 \, \mathrm{mm}$ were drilled into the chopper wheel, enabling the water molecules originating from the ice-dust sample to reach the pressure gauge. To increase the signal-to-noise ratio, the frequency of the rotating chopper wheel was chosen to less than the inverse response time of the pressure gauge. Conversely, high rotational frequencies assure an improved signal stability. Thus, as a compromise, frequencies between $2.0 \, \mathrm{s^{-1}}$ and $3.3 \, \mathrm{s^{-1}}$ were used to meet both requirements. Using a rotating chopper wheel dissects the signal into two parts: in the first case, the rotating chopper wheel reflects all water molecules produced at the surface of the ice-dust sample; hence, only molecules of the surrounding gas are able to hit the sensor; in the second situation, molecules originating from the ice sample's surface and ambient gas molecules can be detected by the pressure gauge. The pressure difference between both situations $\delta p$ was derived by an analysis of the modulated sensor signal with the lock-in amplifier. The quantity $\delta p$ is equal to the pressure increase caused by the sublimating ice-dust sample. The rate of water molecules hitting the sensor $\eta_s$ was calculated assuming elastic collisions between gas molecules and the sensor's surface $A_s$,
\begin{equation}
\eta_{s}(T_i) \, = \, \frac{\delta p}{2 \, m \, \langle v(T_i) \rangle} \, A_s \ \ \ \ \ \mathrm{\left[s^{-1}\right]} \, \mathrm{.}
\label{eq1}
\end{equation}
Here, $m$ is the mass and $\langle v(T_i) \rangle$ the mean velocity of a water molecule, which depends on the temperature of the ice surface $T_i$.
\par
For comparison, the theoretical rate of water molecules hitting the sensor $\eta_t$, which was positioned at the distance $r$ to the sample's surface, can be calculated by
\begin{equation}
\eta_{t}(T_i)\, = \, \alpha(T_i) \, Z(T_i) \, \Omega \, \frac{A_i}{m} \ \ \ \ \ \mathrm{\left[s^{-1}\right]} \, \mathrm{.}
\label{eq2}
\end{equation}
Here, $A_i$ describes the area of the sublimating ice surface and $\Omega = A_s \, / \, (\pi \, r^2)$ is a factor that emerges from the geometry of the experimental setup. The $\pi \, r^2$ term results from the integration over the cosine angular distribution of emitted gas molecules respective to the surface normal of the sample, according to \citet{Knudsen1909}. Due to the geometry of the sample holder, only molecules from the top of the cylindrical sample are able to sublimate. The sublimation rate $Z(T_i)$ of a pure and solid water ice surface is given by the classical Hertz-Knudsen formula \citet{Knudsen1909B}
\begin{equation}
Z(T_i) \, =  \, p_{sub}(T_i) \, \sqrt{\frac{m}{2 \, \pi \, k \, T_i}} \ \ \ \ \ \mathrm{\left[kg \, m^{-2} \, s^{-1}\right]} \, \mathrm{,}
\label{eq3}
\end{equation}
where $k$ is Boltzmann's constant. The sublimation coefficient
\begin{equation}
\alpha_i(T_i) \, = \, \frac{Z'(T_i)}{Z(T_i)}
\label{eq31}
\end{equation}
describes the temperature dependent deviation of the measured sublimation rate $Z'(T_i)$ to the values predicted with the classical Hertz-Knudsen formula \citep{Kossacki1999}. A detailed discussion of the temperature dependency of the sublimation coefficient is given in Sect. \ref{sublimation coefficient}. According to the Clausius-Clapeyron equation and empirical measurements \citep{Fanale1984, Mauersberger2003}, the saturation pressure of water ice, $p_{sub}(T_i)$, can be formulated as
\begin{equation}
p_{sub}(T_i) \, = \, a_1 \, e^{- a_2 / T_i} \ \ \ \ \ \mathrm{\left[Pa\right]} \, \mathrm{.}
\label{eq4}
\end{equation}
Table \ref{SublimationPressureParameter} displays different obtained values for the coefficients $a_1$ and $a_2$ in Eq. \ref{eq4}, including the results of this work (see Sect. \ref{Sublimation of water ice}). Alternative formulations of the temperature dependence of the saturation pressure of water are discussed by \citet{Buck1981}.

\section{Results} \label{Results}
\subsection{Sublimation of hexagonal water ice} \label{Sublimation of water ice}
\begin{figure}
\centering
\includegraphics[angle=180,width=1.00\columnwidth]{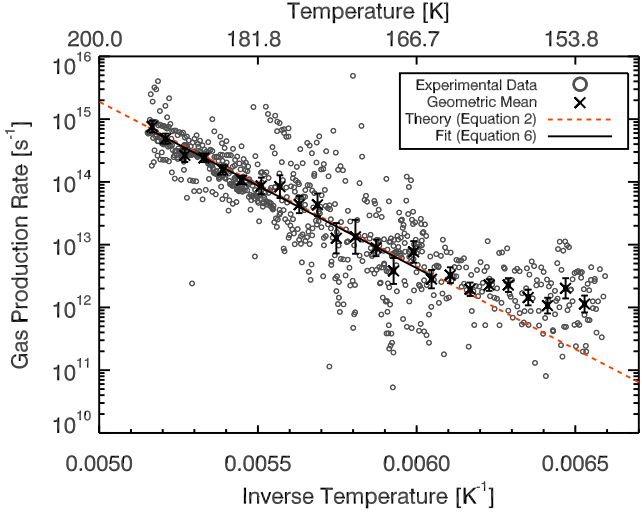}
\caption{Results of the gas production rate measurements of hexagonal water ice (open circles). The solid curve denotes the fit (see Eq. \ref{eq5}) to the geometric means of the binned data (crosses) between $194.1 \, \mathrm{K}$ and $163.9 \, \mathrm{K}$. For this calculation a bin size of $6\times 10^{-5}\,\mathrm{K^{-1}}$ and a sublimation coefficient of $\alpha_i(T_i) = 1$ were used. A comparison with the theoretical gas production rate (dashed curve), given by Eq. \ref{eq2}, shows that the experimental setup is capable to reproduce the known gas production rates of hexagonal water ice over a range of temperatures, which were also measured by \citet{Fanale1984} and \citet{Mauersberger2003}. Note the deviation of the measured gas production rates to the theoretical curve below $163.9 \, \mathrm{K}$, which was caused by the noise of the measuring system.}
\label{ResultsPlot1}
\end{figure}

For the calibration of the sublimation experiment, the gas production rate of hexagonal water ice $\eta_{s}(T_i)$ was measured for different temperatures and compared with the theoretical gas production rate (see Eq. \ref{eq2}). Fig. \ref{ResultsPlot1} shows the results of the measured gas production rates (open circles), as described in Sect. \ref{Experimental}. The experiments were performed in a temperature range between $194.1\,\mathrm{K}$ and $151.7\,\mathrm{K}$. For higher temperatures, the measurement of the gas production rate was not possible, due to the enhanced rate of evaporated water molecules and the associated pressure increase inside the high-vacuum chamber (see Sect. \ref{Sublimation experiment}). Furthermore, the sublimation rate of hexagonal water ice above $\sim194 \, \mathrm{K}$ cannot be explained with the classical Hertz-Knudsen formula, given by Eq. \ref{eq3} (see sublimation coefficient, Sect. \ref{sublimation coefficient}). Thus, the calibration measurements started below this temperature. The lower detection threshold of the gas flux measurement was between $1 \times 10^{12}\,\mathrm{s^{-1}}$ and $3 \times 10^{12}\,\mathrm{s^{-1}}$ and was given by the noise of the measuring system. The collected data were binned in intervals of $6\times 10^{-5}\,\mathrm{K^{-1}}$ and the geometric means of the detected gas fluxes as well as the arithmetic means of the inverse temperatures were computed within these intervals (crosses). The statistical errors of these quantities are denoted by the error bars. For a comparison with the theoretical gas production rate of the ice sample (dashed curve, Eq. \ref{eq2}), the binned data between $194.1 \, \mathrm{K}$ and $163.9 \, \mathrm{K}$ were logarithmically fitted to the function
\begin{equation}
\eta_{i}(T_i)\, = \,  \alpha_i(T_i) \, a_{1,i} \, e^{- a_{2,i} / T_i} \sqrt{\frac{m}{2 \, \pi \, k \, T_i}} \, \Omega \, \frac{A_i}{m} \ \ \ \ \ \mathrm{\left[s^{-1}\right]}
\label{eq5}
\end{equation}
(solid curve), with fitting parameters $a_{1,i}$ and $a_{2,i}$. For these calculations, a sublimation coefficient of $\alpha_i(T_i) \, = \, 1$ was used, because the investigated temperatures were below $\sim194 \, \mathrm{K}$ (see Sect. \ref{sublimation coefficient}). The best agreement between the data and Eq. \ref{eq5} was accomplished for $a_{1,i} = (3.23^{+0.73}_{-0.38}) \times 10^{12} \, \mathrm{Pa}$ and $a_{2,i} =(6134.6 \pm 17.0) \, \mathrm{K}$, which is in good agreement with the results from \citet{Fanale1984} and \citet{Mauersberger2003}. A comparison of the obtained fit parameters with this previous works can be found in Table \ref{SublimationPressureParameter}. These calibration measurements demonstrate that the experimental setup is capable to reproduce known gas production rates of hexagonal water ice within the error tolerance.
\par
The major experimental error was caused by the uncertainty of the temperature sensor's position at the sample's surface. A displacement of $2 \, \mathrm{mm}$ results in a maximal temperature error of $1.6\,\mathrm{K}$, which was comparable to the temperature uncertainty given by the manufacturer. This error can explain the scatter of measurements of the gas production rates.
\par
At temperatures below $163.9 \, \mathrm{K}$, the measured gas flux shows a systematic deviation from the theoretical curve, which was the result of the noise of the measuring system. Thus, we consider a temperature of $163.9 \, \mathrm{K}$ as the lowest temperature accessible to our experimental setup.

\subsection{Gas diffusion through dust layers} \label{Gas diffusion in the porous dust layer}
In total, the gas production rate of twelve ice-dust samples $\eta_{id}(T_i,h)$ (see Sect. \ref{Sample preparation}) were investigated with the sublimation experiment (S0 - S11). The individual measurements are summarized in Table \ref{TableResultsSublimationExperiment}. The height of the dust layer $h$ is given in units of the mean particle diameter.
\begin{table*}[p]
\begin{center}
    \small
    \caption{Summary of the experimental results achieved with the sublimation experiment. The normalized gas production rates of the ice-dust samples $\eta_{id}(T_i,h) / \eta_i(T_i)$ were calculated with Eq. \ref{eq61}. In the experiment $\mathrm{S0}$, a monolayer of glass spheres with a diameter of $2\,\mathrm{mm}$ was positioned on the ice surface (see Sect. \ref{Sample preparation}).}\vspace{1mm}
    \begin{tabular}{lr@{$\pm$}lr@{$\;$}lr@{$\;$}l}
        \hline
        Experiment & \multicolumn{2}{c}{$h$ (Particle Diameter)}  & \multicolumn{2}{c}{$a_{1,id}(h)$ [$\times10^{12}\,\mathrm{Pa}$]} & \multicolumn{2}{c}{$\eta_{id}(T_i,h) / \eta_i(T_i)$}           \\
        \hline
        $\mathrm{S0}^{\dag}$                           & $1.00\,$   & $\,0.00$ & $2.85$ & $^{+0.04}_{-0.04}$ & $0.882$ & $^{+0.034}_{-0.034}$   \\
        S1                                             & $10.62\,$  & $\,2.12$ & $1.30$ & $^{+0.08}_{-0.07}$ & $0.402$ & $^{+0.021}_{-0.020}$    \\
        S2                                             & $17.10\,$  & $\,0.11$ & $1.01$ & $^{+0.08}_{-0.08}$ & $0.313$ & $^{+0.024}_{-0.022}$   \\
        S3                                             & $18.22\,$  & $\,0.34$ & $0.96$ & $^{+0.07}_{-0.07}$ & $0.297$ & $^{+0.020}_{-0.019}$   \\
        S4                                             & $24.59\,$  & $\,0.89$ & $0.76$ & $^{+0.05}_{-0.05}$ & $0.235$ & $^{+0.015}_{-0.014}$    \\
        S5                                             & $26.04\,$  & $\,1.68$ & $0.68$ & $^{+0.11}_{-0.10}$ & $0.211$ & $^{+0.031}_{-0.027}$    \\
        S6                                             & $30.51\,$  & $\,0.11$ & $0.63$ & $^{+0.05}_{-0.05}$ & $0.195$ & $^{+0.015}_{-0.013}$   \\
        S7                                             & $36.55\,$  & $\,1.90$ & $0.56$ & $^{+0.04}_{-0.04}$ & $0.173$ & $^{+0.012}_{-0.012}$   \\
        S8                                             & $47.62\,$  & $\,0.89$ & $0.35$ & $^{+0.05}_{-0.04}$ & $0.108$ & $^{+0.013}_{-0.012}$   \\
        S9                                             & $56.56\,$  & $\,1.12$ & $0.38$ & $^{+0.03}_{-0.02}$ & $0.118$ & $^{+0.007}_{-0.007}$    \\
        S10                                            & $60.13\,$  & $\,1.34$ & $0.27$ & $^{+0.05}_{-0.04}$ &$0.084$ & $^{+0.013}_{-0.011}$   \\
        S11                                            & $67.85\,$  & $\,1.12$ & $0.25$ & $^{+0.03}_{-0.03}$ &$0.077$ & $^{+0.008}_{-0.007}$    \\
        \hline
        \multicolumn{6}{l}{\dag: Investigation of a monolayer of glass spheres instead of the dust layer.}
    \end{tabular}
     \label{TableResultsSublimationExperiment}
\end{center}
\begin{center}
    \small
    \caption{Results of the diffusion experiments. The normalized rates of molecules penetrating through the dust layer $\eta_{df}(h) / \eta_f$ were computed with Eq. \ref{eq9}, using the measured permeabilities of the dust layers $\Theta_d(h)$. In experiment D0, a monolayer of glass spheres was investigated instead of a dust layer. The experiments D2, D10 and D17 were extended to higher pressures to demonstrate that the measurements were not influenced by the increased pressure (see Fig. \ref{PermPressure}).}\vspace{1mm}
    \begin{tabular}{lr@{$\pm$}lr@{$\pm$}lr@{$\pm$}l}
        \hline
        Experiment & \multicolumn{2}{c}{$h$ (Particle Diameter)} & \multicolumn{2}{c}{$\Theta_d(h)$ [$\times 10^{-6} \, \mathrm{m^3 \, s^{-1}}$]} & \multicolumn{2}{c}{$\eta_{df}(h) / \eta_f$}           \\
        \hline
        $\mathrm{D0}^{\dag}$                           & $1.00\,$   & $\,0.00$ & $1514.59\,$ & $\,240.71$ & $0.770\,$  & $\,0.085$    \\
        D1                                             & $12.72\,$  & $\,1.12$ & $86.67\,$ & $\,5.56$ & $0.342\,$  & $\,0.021$    \\
        $\mathrm{D2}^{\ddag}$                          & $13.50\,$  & $\,1.12$ & $63.66\,$ & $\,1.59$ & $0.393\,$  & $\,0.006$    \\
        D3                                             & $14.06\,$  & $\,1.12$ & $41.01\,$ & $\,8.01$ & $0.236\,$  & $\,0.021$    \\
        D4                                             & $19.19\,$  & $\,1.12$ & $47.64\,$ & $\,10.75$ & $0.263\,$  & $\,0.025$    \\
        D5                                             & $20.53\,$  & $\,1.12$ & $58.00\,$ & $\,3.24$ & $0.258\,$  & $\,0.021$    \\
        D6                                             & $24.55\,$  & $\,1.12$ & $26.92\,$ & $\,4.11$ & $0.215\,$  & $\,0.021$    \\
        D7                                             & $27.90\,$  & $\,1.12$ & $40.52\,$ & $\,4.47$ & $0.195\,$  & $\,0.007$    \\
        D8                                             & $28.79\,$  & $\,1.12$ & $42.59\,$ & $\,1.87$ & $0.204\,$  & $\,0.021$    \\
        D9                                             & $32.36\,$  & $\,1.12$ & $22.58\,$ & $\,2.56$ & $0.187\,$  & $\,0.015$    \\
        $\mathrm{D10}^{\ddag}$                         & $34.70\,$  & $\,1.12$ & $21.22\,$ & $\,0.65$ & $0.177\,$  & $\,0.004$    \\
        D11                                            & $35.04\,$  & $\,1.12$ & $34.78\,$ & $\,1.24$ & $0.173\,$  & $\,0.016$    \\
        D12                                            & $41.29\,$  & $\,1.12$ & $21.99\,$ & $\,1.51$ & $0.116\,$  & $\,0.008$    \\
        D13                                            & $44.19\,$  & $\,1.12$ & $18.88\,$ & $\,1.29$ & $0.131\,$  & $\,0.011$    \\
        D14                                            & $50.43\,$  & $\,1.12$ & $19.29\,$ & $\,1.92$ & $0.133\,$  & $\,0.005$    \\
        D15                                            & $54.45\,$  & $\,1.12$ & $10.50\,$ & $\,1.05$ & $0.077\,$  & $\,0.001$    \\
        D16                                            & $57.58\,$  & $\,1.12$ & $13.55\,$ & $\,1.06$ & $0.098\,$  & $\,0.004$    \\
        $\mathrm{D17}^{\ddag}$                         & $72.75\,$  & $\,1.12$ & $10.42\,$ & $\,0.51$ & $0.096\,$  & $\,0.004$    \\
        \hline
        \multicolumn{6}{l}{\dag: Investigation of a monolayer of glass spheres instead of the dust layer.}\\
        \multicolumn{6}{l}{\ddag: Measurements extended to higher pressures.}
    \end{tabular}
     \label{TableResultsDiffusionExperiment}
\end{center}
\end{table*}
\par
\begin{figure}
\centering
\includegraphics[angle=180,width=1.00\columnwidth]{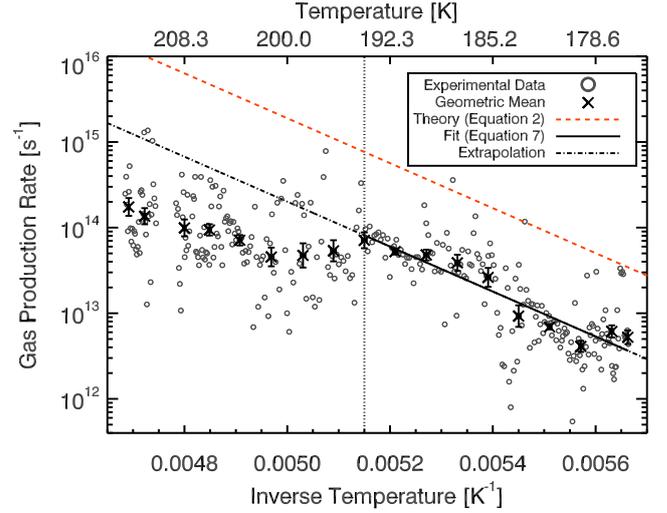}
\caption{Example of the determination of the normalized gas production rate of an ice-dust sample in the sublimation experiment. The open circles display the measured gas production rates of experiment S9, for which the height of the dust layer was $(56.56\pm1.12)$ particle diameters. Fitting the geometric means of the data (crosses) below $194.2 \, \mathrm{K}$ with Eq. \ref{eq6} (solid curve) and comparing this result with the gas production rate of hexagonal water ice without an additional dust layer (see Eq. \ref{eq5}) gives the normalized gas production rate of the ice-dust sample. For comparison, the theoretical gas production rate of a pure ice sample is visualized by the dashed curve (see Eq. \ref{eq2}). The best match to the data was achieved for $a_{1,id}=(0.38^{+0.03}_{-0.02})\times 10^{12}\,\mathrm{Pa}$, which results in a normalized gas production rate of $0.118^{+0.007}_{-0.007}$. An extrapolation of the fit to higher temperatures (dash dotted curve) demonstrates the strong variation of the sublimation coefficient above $\sim194 \, \mathrm{K}$ (dotted line). For this data analysis a bin size of $6\times10^{-5}\,\mathrm{K^{-1}}$ was used.}
\label{S9}
\end{figure}

From the deviation of the measured gas production rates to the values for pure hexagonal water ice without a dust layer (see Sect. \ref{Sublimation of water ice}), the normalized gas production rates of the ice-dust samples, $\eta_{id}(T_i,h) / \eta_i(T_i)$, were calculated. For an explanation of this derivations, the results of the experiment S9 are shown in Fig. \ref{S9}.
\par
In this experiment a dust layer with a height of $(56.56\pm1.12)$ particle diameters was used. The measured gas production rates (open circles) were binned in intervals of $6\times10^{-5}\,\mathrm{K^{-1}}$ and the geometric means of the gas production rate as well as the arithmetic means of the inverse temperatures within these intervals were calculated (crosses). The error bars denote the statistical errors of the derived values. To compare the gas production rates of the ice-dust sample with the gas flux of a pure (i.e. dust-free) hexagonal ice samples (dashed curve, Eq. \ref{eq2}), the following function was used to fit the data,
\begin{equation}
\eta_{id}(T_i,h)\, = \, \alpha(T_i) \, a_{1,id}(h) \, e^{- a_{2,id} / T_i} \sqrt{\frac{m}{2 \, \pi \, k \, T_i}} \, \Omega \, \frac{A_i}{m} \ \mathrm{\left[s^{-1}\right]}
\label{eq6}
\end{equation}
(solid curve in Fig. \ref{S9}). An additional dust layer on top of the ice surface decreases the gas production rate of the ice sample. However, this reduction should not depend on the temperature of the sample, if we assume that the gas molecules are only scattered by the individual particles of the dust layer. If the gas molecules were absorbed on the surface of the dust particles, evaporation of the trapped gas molecules occurs over time. In this case, the rate of evaporation of the condensed water molecules inside the dust layer is bigger than the rate of water molecules supplied by the ice surface, if the temperature of the dust layer is higher than the temperature of the ice surface. Thus, the dust layer loses the water ice after some period of time. Even in this case, the influence of the dust layer on the gas production rate of the ice sample is still not affected by the temperature of the ice surface (a detailed discussion of the thermal influence of the dust layer on the gas production rate is given at the end of this Section). Thus, the temperature of the dust layer and, therewith, the temperature of the ice surface will not affect the influence of the dust layer on the gas production rate. Thus, we assume that the dust layer on top of the ice sample does not influences the slope of the function in Eq. \ref{eq6}. Therefore, the parameter $a_{2,id}$, which determines the slope of the fit was fixed to fit the data, i.e. $a_{2,id} = a_{2,i}$ (see Sect. \ref{Sublimation of water ice}). Thus, the coefficient $a_{1,id}(h)$ was the only free parameter, which was varied to fit the binned gas production rates with Eq. \ref{eq6}. From the results of the fits, the normalized gas production rates (rates of molecules penetrating through the dust layer) were calculated,
\begin{equation}
\frac{\eta_{id}(T_i,h)}{\eta_i(T_i)} \, = \, \frac{a_{1,id}(h)}{a_{1,i}} \, \mathrm{.}
\label{eq61}
\end{equation}
For the S9 experiment, the best match to the binned data was found for $a_{1,id}=(0.38^{+0.03}_{-0.02})\times 10^{12}\,\mathrm{Pa}$. This results in a normalized gas production rate of $0.118^{+0.007}_{-0.007}$.
\par
The fit to the data could only be realized below a critical temperature of $194.2 \, \mathrm{K}$ (solid line), due to a change in the sublimation coefficient at higher temperatures (see Sect. \ref{sublimation coefficient}). To demonstrate the decrease of the sublimation coefficient at higher temperatures, the fit to the binned data was extrapolated (dash dotted line).
\par
\begin{figure}
\centering
\includegraphics[angle=180,width=1.00\columnwidth]{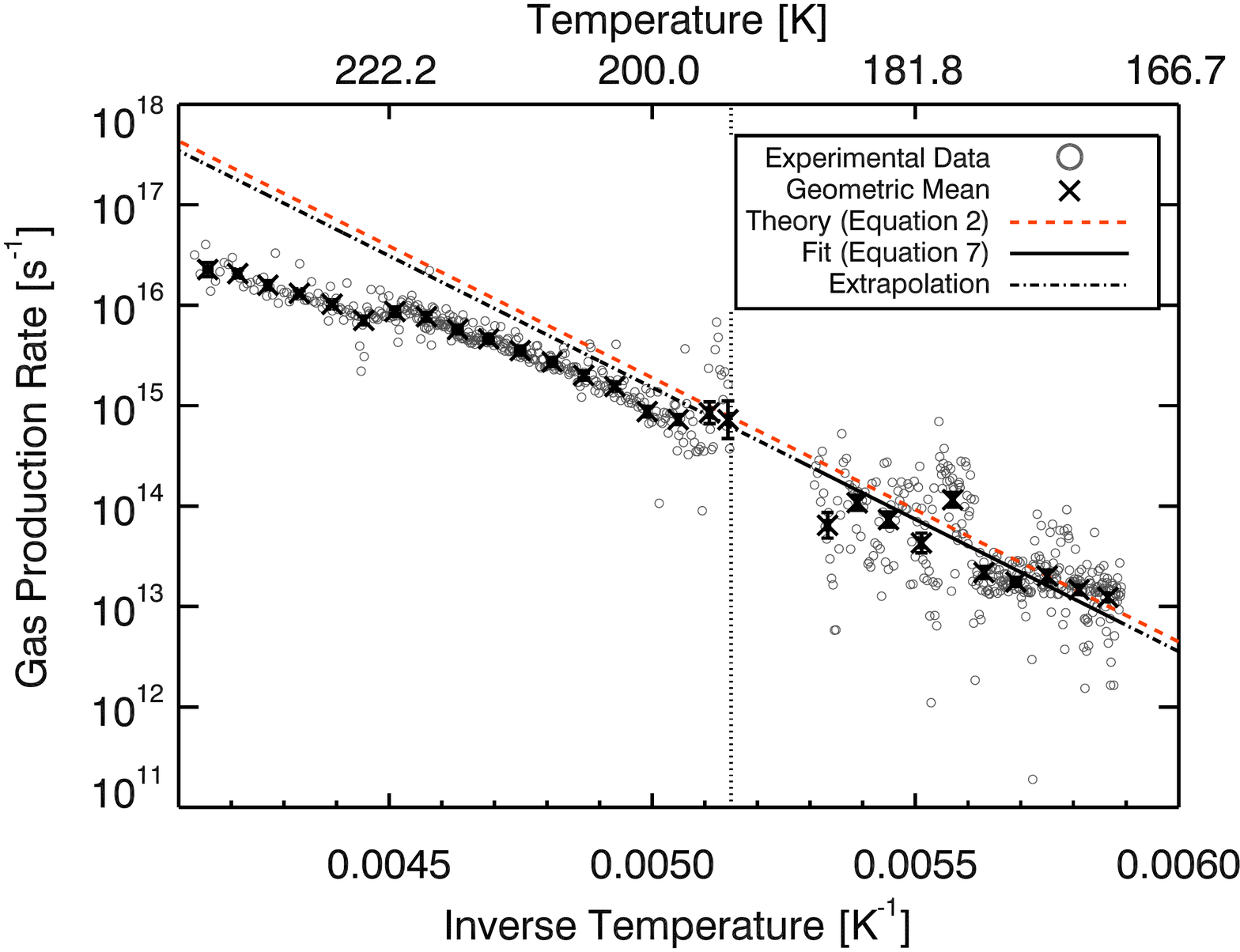}
\caption{Results of the experiment S0, in which 133 monodisperse glass spheres were positioned on the ice sample to investigate the normalized gas flux through a monolayer of dust. The achieved data (circles) were binned in intervals of $6 \times 10^{-5} \, \mathrm{K^{-1}}$ and the geometric means of the gas fluxes and the arithmetic means of the inverse temperatures were derived (crosses). The statistical errors of theses quantities are visualized by the error bars. The best fit to the data using Eq. \ref{eq6} was achieved for $a_{1,id}=(2.85^{+0.04}_{-0.04}) \times 10^{12} \, \mathrm{Pa}$ (solid curve). This results in a normalized gas flux of $0.882^{+0.034}_{-0.034}$ for the monolayer dust. Note the deviation of the detected gas fluxes to the extrapolated fit (dash dotted curve) at temperatures above $194.2 \, \mathrm{K}$.}
\label{MonolayerSublimationExperiment}
\end{figure}

The obtained fit parameters and the derived normalized gas production rates for the different sublimation experiments are presented in Table \ref{TableResultsSublimationExperiment}. For the S0 experiment, a monolayer of 133 glass spheres was added on top of the ice surface (see Sect. \ref{Sample preparation}). The result of this experiment is visualized in Fig. \ref{MonolayerSublimationExperiment}. Note the deviation of the measured gas production rates to the extrapolated fit (dash dotted curve) above $\sim194 \, \mathrm{K}$, due to the decrease of the sublimation coefficient at higher temperatures (see Sect. \ref{sublimation coefficient}).
\par
Further investigations of the gas diffusion through dust layers at room temperature were carried out using the diffusion experiment. In 18 experiments (D0 - D17), the permeability of the dust layer and the normalized rate of molecules penetrating through the dust layer were measured at room temperature. The results of these experiments are summarized in Table \ref{TableResultsDiffusionExperiment}.
\par
The Knudsen number, defined by
\begin{equation}
Kn \, = \, \frac{\lambda}{d_{eff}}  \, \mathrm{,}
\label{y1}
\end{equation}
with the mean free path of the gas molecules, $\lambda$, and the effective pore size of the material, $d_{eff}$, is characteristic for the flow regime. Free molecular flow is expected for $Kn > 1$. Due to the low gas pressures in our experiments, the Knudsen numbers were always $Kn > 3.3$ so that collisions of gas molecules with the particular medium dominate over intermolecular collisions, and the mass is transferred due to the pressure gradient according to Knudsen's law, where gas molecules migrate independently of each other. The diffusive flux of an ideal gas is described by Fick's first law,
\begin{equation}
j_{mol} \, = \, D_{Kn} \, \frac{dC}{dx} \ \ \ \ \ \mathrm{\left[ mol \, m^{-2} \, s^{-1} \right]} \, \mathrm{,}
\label{y2}
\end{equation}
relating the diffusion flux $j_{mol}$ to the gradient of the molecular concentration $C$ $\left[\mathrm{mol \, m^{-3} }\right]$. $D_{Kn}$ $\left[\mathrm{m^2 \, s^{-1}}\right]$ is the Knudsen diffusion coefficient or diffusivity. With the ideal gas law, Fick's first law can be formulated in terms of the pressure gradient,
\begin{equation}
j_{mol} \, = \, \frac{D_{Kn}}{R \, T_g} \, \frac{dP}{dx} \, \mathrm{,}
\label{y3}
\end{equation}
where $R$ is the ideal gas constant and $T_g$ the temperature of the gas. This equation can be rewritten for the flux $j$ measured in numbers of molecules flowing through a unit area per time interval,
\begin{equation}
j \, = \, \frac{D_{Kn}}{k \, T_g} \, \frac{dP}{dx} \ \ \ \ \ \mathrm{\left[m^{-2} \, s^{-1} \right]}\, \mathrm{.}
\label{y4}
\end{equation}
For the steady gas flow through a random porous medium, the effective Knudsen diffusivity can be introduced by
\begin{equation}
D_{Kn} \, = \, - \frac{1}{3} \, d_{eff} \, \frac{\psi}{\tau} \, \sqrt{\frac{8 \, k \, T_g}{\pi \, m}} \, \mathrm{,}
\label{y5}
\end{equation}
where $\psi$ is the transport porosity of the medium and $\tau$ is the tortuosity factor. The medium used in the diffusion experiment has only open pores. Thus, the transport porosity equals the total porosity. The tortuosity factor is defined as the square of the ratio of the effective path (or actual length $L_e$) to the thickness of the porous layer $h$, or as the square of the ratio of the average free path traveled by the species $\langle l\rangle$ to the average displacement in the direction of diffusion $\langle x\rangle$ \citep{Boudreau1996}
\begin{equation}
\tau \, = \, \frac{L_e^2}{h^2} \, = \, \frac{\langle l^2\rangle}{\langle x^2\rangle} \, \mathrm{.}
\label{y6}
\end{equation}
From the measured pressure difference $\Delta p$ and the adjusted gas flux through the dust layer and the filter paper $\phi_{df}$, the permeability of the dust layer $\Theta_d(h)$ was determined by
\begin{equation}
\Theta_d(h) \, = \, \left(\frac{\Delta p}{\phi_{df}} - \frac{1}{\Theta_f}\right)^{-1} \ \ \ \ \ \mathrm{\left[m^3 \, s^{-1}\right]} \, \mathrm{.}
\label{eq7}
\end{equation}
Here, $\Theta_f$ is the permeability of the filter paper, which was estimated in several calibration experiments. This inter-relation arises from the fact that the permeability is given by the inverse resistivity to the gas flow through the material. This resistivity can be derived from the quotient of the pressure difference (potential difference) and the gas flux (current). Thus, the dust layer acts like a resistor to the gas flux. In the case of the sublimation experiment, the ice surface of the sample possesses a certain internal resistance, which is given by the inverse sublimation probability of the water molecules. Furthermore, the internal resistance in case of the diffusion experiment is caused by the filter paper. Therefore, the resistivity of the dust layer together with the internal resistivity can be described with two resistors in series-connection (resistivity model).
\par
\begin{figure}[t]
\centering
\includegraphics[angle=180,width=1.00\columnwidth]{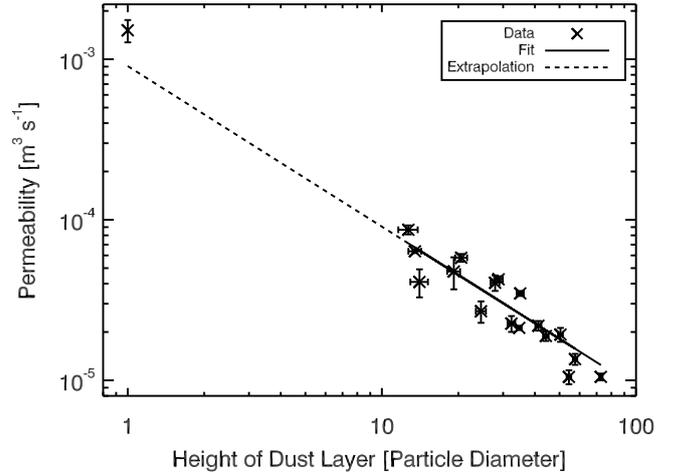}
\caption{Derived permeabilities for different heights of the dust layer, calculated with Eq. \ref{eq7} (crosses) and fitted with Eq. \ref{eq8} (solid curve). The result of the D0 experiment was not included into the fit, to achieve a better match to the data between 12.72 and 72.75 particle diameters. From the fit, the permeability coefficient $\Theta_0 =  (4.96 \pm 0.98) \times 10^{-8}\,\mathrm{m^4 \, s^{-1}}$ was derived.}
\label{AbsPerm}
\end{figure}

Fig. \ref{AbsPerm} shows the derived permeabilities as a function of the height of the dust layer (crosses). From the fit to these data (solid curve), the following correlation between the permeability and the height of the dust layer was found,
\begin{equation}
\Theta_d(h) \, = \, \frac{\Theta_0}{h}\, \mathrm{,}
\label{eq8}
\end{equation}
where $\Theta_0 =  (4.96 \pm 0.98) \times 10^{-8}\,\mathrm{m^4 \, s^{-1}}$ is the permeability coefficient of the dust layer. The result of the D0 experiment was not included into the fit, to achieve a better match to the derived permeabilities in the interval between 12.72 and 72.75 particle diameters. The dashed curve demonstrates the extrapolation of the applied fit. The major error in these measurements are caused by the uncertainty of the pressure signal and the gas flux adjustment.
\par
A comparison between Eqs. \ref{y4}, \ref{eq7}, and \ref{eq8} shows that the effective Knudsen diffusivity is connected with the permeability coefficient by the evident ratio
\begin{equation}
D_{Kn}\, = \, \frac{\Theta_0}{A_d} \, \mathrm{,}
\label{y6}
\end{equation}
where $A_d$ is the cross section of the sample. Because the gas temperature and porosity of the sample are known, one can evaluate the ratio of the effective pore size to the tortuosity factor. Further evaluation requires the use of a computer model describing molecular diffusion in close packed porous media. The construction of such a model is in progress and corresponding quantitative analysis of the results will be presented in a forthcoming paper.
\par
\begin{figure}[t]
\centering
\includegraphics[angle=180,width=1.00\columnwidth]{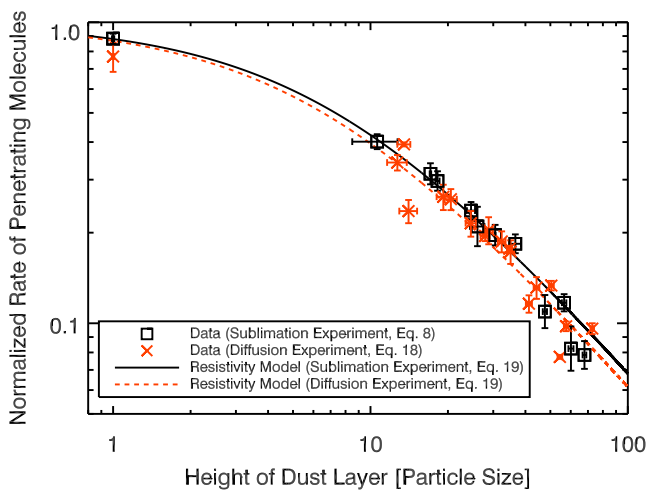}
\caption{Derived normalized rates of molecules penetrating through the dust layer, measured with the sublimation experiment (squares) and the diffusion experiment (crosses) as a function of the height of the dust layer. The data were fitted with Eq. \ref{eq10} (resistivity model). The best match to the data was achieved with $b_{id} = (7.31 \pm 0.20)$ particle diameters for the sublimation experiment (solid curve) and $b_{df} = (6.54 \pm 0.38)$ particle diameters for the diffusion experiment (dashed curve). The height of the dust layer is given in units of the monomer diameter. Mind that two different monomer sizes were used in the experiments, i. e. $2\,\mathrm{mm}$ (monolayer experiment) and $(54.7\pm11.0)\,\mathrm{\mu m}$ (any other experiment).}
\label{PermeabilityPlot}
\end{figure}
To compare the results of the diffusion experiment with the results of the sublimation experiment, the normalized rates of molecules penetrating through the dust layer and the filter paper $\eta_{df}(h) / \eta_f$ were calculated using the resistivity model
\begin{equation}
\frac{\eta_{df}(h)}{\eta_f} \, = \, \frac{\Theta_d(h)}{\Theta_f + \Theta_d(h)}  \, \mathrm{.}
\label{eq9}
\end{equation}
Fig. \ref{PermeabilityPlot} presents the comparison between the normalized gas production rates of the ice-dust samples (squares, Eq. \ref{eq61}) and the resulting normalized rates of molecules penetrating through the (warm) dust layer and the filter paper (crosses, Eq. \ref{eq9}). The error bars indicate the statistical errors, which were derived from the errors of the fits (sublimation experiment) and from the standard deviations to the arithmetic means (diffusion experiment). The decrease of the normalized diffusion rates for increasing heights of the dust layer were fitted using the resistivity model
\begin{equation}
\frac{\eta_{id,df}(h)}{\eta_{i,f}} \, = \,  \frac{1}{1 + b^{-1} h} \, \mathrm{.}
\label{eq10}
\end{equation}
Here, $b$ is the height in particle diameters at which $50\,\%$ of the initial gas flux penetrates through the dust layer. The best match to the normalized diffusion rates was found for $b_{id} = (7.31 \pm 0.20)$ particle diameters in case of the ice-dust samples (solid curve) and $b_{df} = (6.54 \pm 0.38)$ particle diameters for the dust layers at room temperature. Both data sets show the same functional behavior with respect to the height of the dust layers. This accordance can have two different causes:
\begin{enumerate}
\item Due to the low heat conductivity of the loose material \citep{Krause2011}, the temperature decrease rate inside the dust layer was much smaller than the cooling rate of the ice. Therefore, the temperature of the bulk of the dust layer remained high enough so that no condensation and recondensation of water molecules occurred inside the dust layer. A comparison of the heat conductivity of hexagonal water ice \citep[$3.26\,\mathrm{W\,m^{-1}\,K^{-1}}$;][]{Slack1980} with the heat conductivity of relatively densely packed silica particles \citep[$\sim10^{-2}\,\mathrm{W\,m^{-1}\,K^{-1}}$;][]{Krause2011} demonstrates that the temperature gradient inside the dust layer should be much higher than the temperature gradient within the ice sample. Thus, only the first few layers of the dust particles above the ice surface can remain cold enough for condensation of water molecules.
\item If we assume that part of the dust layer was principally cold enough for ice condensation to take place, the evaporation rate of water molecules from inside the dust layer must be, however, higher than the evaporation rate of the icy surface, due to the higher temperature of the dust layer. Thus, the dust layer cannot quantitatively store water molecules in form of icy condensates on the dust particles' surfaces.
\end{enumerate}
\par
In our experiments we measured the surface temperature of the ice surface, therefore, the thermal influence of the presence or absence of the dust layer is unimportant for the interpretation of the gas production rate of the ice surface. The temperature of the ice surface is influenced by back-scattered molecules inside the dust layer and by the transported heat through the dust layer and through the ice sample, but the gas production rate is determined by the actual surface temperature, which was continuously measured. However, in future experiments we will concentrate on the heat transport through the dust layers by insolation. The influence of the grain size and the porosity on the heat transport through dust layers will also be studied.
\par
The match between the two experiments also implies that the diverse pressure regimes of the two different experimental environments have not affected the measurements. In both experiments, the mean free path of the gas molecules was larger than the mean pore size of the material, which was of the same order as the mean particle size. A calculation of the Knudsen numbers $Kn$ demonstrates that the experiments were performed in the free molecular flow regime, i.e. $Kn = 9877.5$ (sublimation experiment) and $Kn = 3.3$ (diffusion experiment), respectively. Thus, the obtained results should not be influenced by the different pressure regimes as long as the Knudsen numbers guarantee a free molecular flow through the dust layer.
\par
\begin{figure}[t]
\centering
\includegraphics[angle=180,width=1.00\columnwidth]{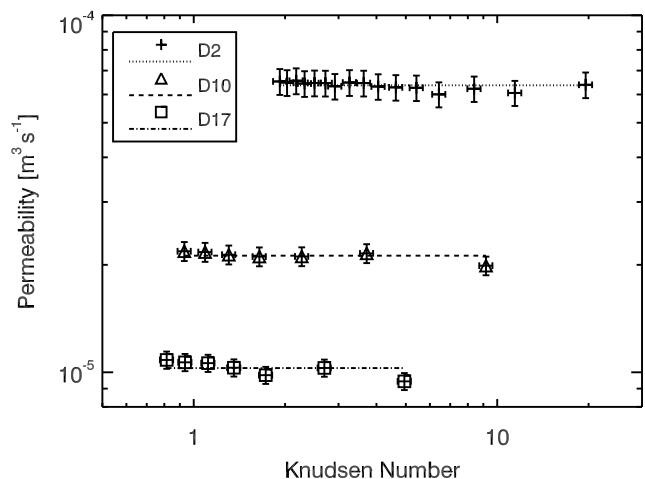}
\caption{Extended permeability measurements for lower Knudsen numbers performed with experiments D2 (pluses), D10 (triangles) and D17 (squares). The means of the individual measurements are denoted by the dotted, dashed and dash dotted lines.}
\label{PermPressure}
\end{figure}

Further investigations with the diffusion experiment at higher pressures (D2, D10, D17) have shown that the permeability remained constant even for smaller Knudsen numbers (see Fig. \ref{PermPressure}). For the calculation of the Knudsen numbers, the mean particle diameter was used to determine the characteristic length of the system. Furthermore, the presence of a laminar flow through the dust during the diffusion experiments was confirmed by an analysis of the Reynolds number, which was in the range between 0.8 and 9.8. The gas flow through the dust layer on top of the ice sample was obviously not influenced by turbulence.

\subsection{Sublimation coefficient} \label{sublimation coefficient}
The sublimation coefficient (see Eq. \ref{eq31}) is introduced to describe the temperature dependent deviation of the measured sublimation rate to the expected theoretical value of hexagonal water ice, given by the classical Hertz-Knudsen formula (see Eq. \ref{eq3} and Fig. \ref{MonolayerSublimationExperiment}). From the standpoint of statistical physics, the sublimation of water ice should be considered within the general theory describing adsorption and desorption processes. In the case when the adsorbate and adsorbent are identical and the vapor pressure is negligible in comparison with the saturation pressure determined from the Langmuir adsorption isotherm, the equation describing the sublimation of the ice crystal is reduced to Eq. \ref{eq3}, which is widely used in planetary physics. It is common knowledge that this formula is trivially derived from the kinetic theory of gas using just one assumption about the equilibrium distribution of molecular velocity  \citep{SkorovRickmann1995}. This means that all information about the microphysical processes determining the actual sublimation rate as well as the growth rate of the ice crystal is hidden (masked) in the so-called sublimation coefficient, which obviously has a complex physical background. A detailed microphysical investigation of this coefficient is indeed a sophisticated theoretical task, which is far beyond the scope of this work. It should be only mentioned that the rate of sublimation is influenced by the shape of the crystal, the structure of its surface, the presence of impurities and other physical factors. For example, the theoretical study of the sublimation rate for the case of an ideal homogeneous surface was presented by \citet{Shulman1972} where he treated the sublimation as a variant of a more general process of adsorption-desorption using the method of chemical potentials. A more complete consideration of the evaporation problem was carried out in the work performed by \citet{KnakeStranskii1959} in which the authors took a set of parameters into account characterizing the heterogeneity of the crystal, i.e. a variable coordination number and, therefore, variable activation energy of the molecules, the geometry of the crystal lattice, and the distance between the defects. It should be emphasized that it is the complexity of the problem, which makes the theoretical formula unsuitable for practical use, because many of these microscopic characteristics are not well defined in laboratory experiments, and in addition, they can vary over time.
\par
This is why new direct experimental measurements of crystalline-ice sublimation/growth under particular non-equilibrium conditions (mainly given by the temperature and pressure range) are so important for planetary applications. The assumption that sublimation is the exact reverse of a growth process allows us to use experimental data obtained during the investigation of the latter \citep{PruppacherKlett1978}. The corresponding set of references can be found in the work of \citet{Kossacki1999} and \citet{Nelson1997}. It should be noted that the deviation of the sublimation coefficient from unity were observed by \citet{KramersStemerding1951}, \citet{IsonoIwai1969}, \citet{Beckmann1982}, and later by \citet{Kossacki1999} with quite different experimental setups. At the same time, this effect was clearly observed at about the same temperatures, i.e. above $~190 \mathrm{K}$. At lower temperatures, there is a good agreement between the theoretical estimate from the Hertz-Knudsen formula (see Eq. \ref{eq3}) and the achieved experimental results. Hereafter, we present the results of our systematic evaluation of the sublimation coefficient carried out under experimental conditions reasonable for the planetary science applications.
\par
\begin{figure}[t]
\centering
\includegraphics[angle=180,width=1.00\columnwidth]{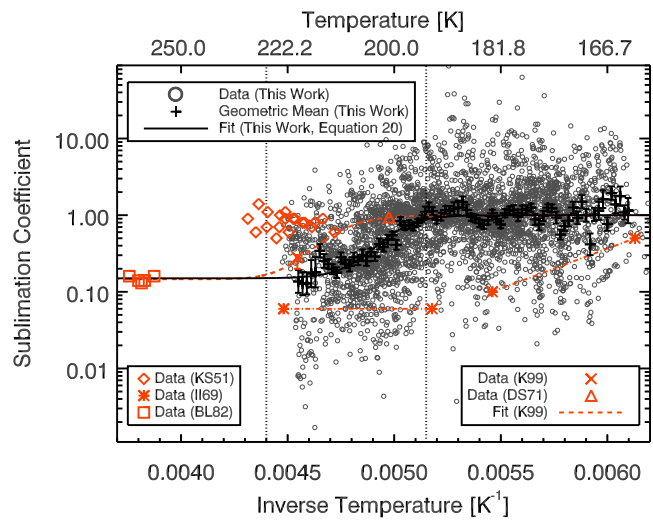}
\caption{Temperature dependence of the sublimation coefficient of hexagonal water ice, calculated from the gas production rates measured with the sublimation experiment (experiments: S0 - S11). The retrieved data (open circles) were binned in intervals of $10^{-5}\,\mathrm{K^{-1}}$, and the geometric means of the derived sublimation coefficients as well as the arithmetic means of the inverse temperatures were computed (pluses). According to these results, a new function for the temperature dependence of the sublimation coefficient is proposed (solid curve, see Eq. \ref{eq11}). A good match to the data was retrieved for $c_1 = (11.58 \pm 0.24) \times 10^{3} \, \mathrm{K}$ and $c_2 = (4.99 \pm 0.20)\times10^{-3} \, \mathrm{K^{-1}}$. For comparison, the results of the following works are also visualized: \citet[diamonds]{KramersStemerding1951}, \citet[asterisks, red dash dotted curve]{IsonoIwai1969}, \citet[squares]{Beckmann1982}, \citet[triangle]{Davy1971} and \citet[cross, dashed curve]{Kossacki1999}. The red dash dotted lines denote the expected temperature range of the sublimation coefficient measured by \citet{IsonoIwai1969}.}
\label{Alphaplot}
\end{figure}
The dashed curve in Fig. \ref{Alphaplot} visualizes the temperature dependence of the sublimation coefficient, derived by \citet{Kossacki1999}. This retrieved functional behavior was only based on three different experimental works (\citet[squares]{Beckmann1982}, \citet[cross]{Kossacki1999}, and \citet[triangle]{Davy1971}), performed in three limited temperature intervals. For further comparison, the results of previous works are also presented in Fig. \ref{Alphaplot}: \citet[diamonds]{KramersStemerding1951} and \citet[asterisks]{IsonoIwai1969}. However, the disagreement between these different experimental works is evident. Thus, a more detailed investigation of the temperature dependence of the sublimation coefficient in a broader temperature range and under well defined experimental conditions was conducted, using the sublimation experiment.
\par
Using the results of the experiments S0 - S11, the sublimation coefficient was derived from the deviation of the measured gas production rates to the applied fits (see Eq. \ref{eq6}, sublimation experiment) and the related extrapolations (see Fig. \ref{S9}). Fig. \ref{Alphaplot} presents the calculated sublimation coefficients (open circles). The data were binned in intervals of $10^{-5}\,\mathrm{K^{-1}}$ and the geometric means of the derived sublimation coefficients as well as the arithmetic means of the inverse temperatures were computed (pluses). This calculations were only conducted in intervals, where three or more individual measurements were performed. Due to the retrieved results, a new formulation for the temperature dependence of the sublimation coefficient is proposed,
\begin{equation}
\alpha(T_i) \, = \, \frac{0.854}{1 + \mathrm{exp}\big[-c_1 \, \big(T_i^{-1} - c_2\big)\big]} + 0.146
\label{eq11}
\end{equation}
(solid curve), with $c_1 = (11.58 \pm 0.24) \times 10^{3} \, \mathrm{K}$ and $c_2 = (4.99 \pm 0.20)\times10^{-3} \, \mathrm{K^{-1}}$. This new function is in good agreement with the results from \citet{Kossacki1999} for temperatures above $\sim 227 \, \mathrm{K}$ and below $\sim 194 \, \mathrm{K}$. Between these two temperatures, the new formulation deviates from the functional behavior published by \citet{Kossacki1999} (dashed line). However, the new proposed temperature dependence of the sublimation coefficient is based on $4,048$ data points, which were achieved by eleven individual measurements.
\par
The different formulations of the temperature dependencies of the sublimation coefficient have a huge influence on the energy balance of sublimating ice surfaces and thus on the thermal modeling of icy bodies in the solar system.

\section{Conclusions and implications}
The overt lack of experimental data on ice sublimation under non-equilibrium conditions (low pressure) and on transport properties (i.e. on the absolute permeability) of porous media results in a lack of understanding of cometary activity. Therefore, we focused on two principal questions in this paper: (1) how can sublimation be kept alive beneath the covering porous dust layer and (2) how does the sublimation rate depend on the temperature of the sublimating ice?
\par
We found that the normalized rate of penetrating molecules through a dust layer is inversely proportional to its thickness. This result can be explained with the introduced resistivity model (see Sect. \ref{Gas diffusion in the porous dust layer}), in which the dust layer is treated as a resistor to the gas flux. Using this model, we are able to calculate the thickness at which $50\,\%$ of the initial gas flux penetrates through the dust layer: $(7.31 \pm 0.20)$ particle diameters (sublimation experiment) and $(6.54 \pm 0.38)$ particle diameters (diffusion experiment). This accordance among the two different experiments may have two different causes. Either the temperature decrease rate inside the dust layer was much smaller than the cooling rate of the ice, due to the low heat conductivity of the loose material \citep{Krause2011}. Therefore, the temperature of the dust layer probably remained high enough so that no condensation and recondensation of water molecules have occurred inside the dust layer. Or, the temperature of part of the dust layer was principally cold enough for ice condensation to take place. In this case, the evaporation rate of water molecules from inside the dust layer must be, however, higher than the evaporation rate of the icy surface, due to the higher temperature of the dust layer. Thus, the dust layer cannot quantitatively store water molecules in form of icy condensates on the dust particles' surfaces.
\par
The second objective of this work was the systematic evaluation of the so-called sublimation coefficient, defined as the ratio between experimental and maximal sublimation rate (Eq. \ref{eq31}). Therefore, we demonstrated that the experimental setup is capable to reproduce the known gas production rates of hexagonal water ice. From the deviation to the theory, the sublimation coefficient was calculated for temperatures ranging from $223.4 \,\mathrm{K}$ to $157.4\,\mathrm{K}$. This investigation revealed that the sublimation coefficient is unity for temperatures below $\sim 194 \, \mathrm{K}$. Between $\sim 194 \, \mathrm{K}$ and $\sim 227 \, \mathrm{K}$, the sublimation coefficient decreases by approximately one order of magnitude. This behavior has a tremendous influence on the energy balance of sublimating ice surfaces and, therefore, on the thermal modeling of icy bodies in the solar system.
\par
Until recently, the questions which are the subject of this work had only an indirect relation to the physics of comets, because the existence of a non-volatile porous crust on the surface of the cometary nucleus as well as the sublimation of volatiles through this crust remained unproven hypotheses and theoretical assumptions. However, the situation has changed dramatically over the past few years, when successful space missions (Deep Space I, Stardust, Deep Impact, EPOXI) delivered a lot of new facts relating to the physics of cometary nuclei.
\par
Today, we possess high-resolution images of five nuclei of periodic comets. These images, together with spectral and infrared observations, have given us solid evidence that all cometary nuclei observed till now are largely covered by a porous, dark, non-volatile crust. Especially exciting results were obtained by the Deep Impact mission to comet Tempel 1 and later to comet Hartley 2. The thermal inertia of the surface of comet Tempel I was found to be very low \citep{Groussin2007}. This result indicates that a high-porosity material forms the surface layer \citep{Krause2011}. At the same time, only traces of water ice were found on the surface \citep{Sunshine2007}. These findings, together with the measured gas activity \citep{Schleicher2006} can be considered as a strong argument in support of ice sublimation beneath a thin dust layer. The latest images of comet Hartley 2, taken during the flyby of the Deep Impact spacecraft (http://epoxi.umd.edu), give evidence of the high activity of the nucleus, with numerous jets visible on the surface. The jets are still alive on the night side of the nucleus. As in the case of other comets, the surface of Hartley 2 is dark, an indication that it is covered by dust. Thus, one can expect that the transport of sublimation products through a porous non-volatile dust layer is an important process on comet Hartley 2. All these data make the investigation of mass transport through porous media an interesting and important endeavor.
\par
Recently, we discussed in detail the theoretical aspect of the first question in application to the cometary physics \citep{Skorov2010}. In that paper, we reviewed the used approaches and presented a microphysical computational model of Knudsen diffusion in random porous media (RPM) formed by packed monodisperse spheres. The main transport characteristics, such as the mean free path distribution and the relative permeability of the porous slab, were calculated for a high-porosity medium (porosity was above $65\,\%$) and compared with that obtained by classical capillary models. Both approaches (RPM and capillary models) predict that, in the stationary stage of diffusion, the relative permeability of the porous layer is inversely proportional to its thickness. This relation was observed also in all our experiments. At the same time, a clear difference between theoretical and experimental data obtained for the thin layer has been detected. For example, we got an experimental permeability of a monolayer of dust that is significantly higher than the theoretical predictions. We also note that the absolute weakening of the gas flux was smaller in the experiments. This lack of correspondence between model and experiment may be caused by an ill-determination of the boundary condition in the theoretical models, when the molecules scattered back were excluded from further consideration.  Obviously, new laboratory and computer experiments are necessary to clarify the situation and to validate the new, more sophisticated theoretical models of cometary nuclei.
\par
The second focus of this research is the evaluation of the sublimation coefficient. Although the variability of this characteristics is confirmed experimentally (see Sect. \ref{sublimation coefficient}), the effect has not received the attention it deserves in cometary research. The publication by \citet{Kossacki1999} remains one of a few attempts to explore the consequences of this phenomenon for the transport processes in cometary nuclei. It was shown, for example, that for small heliocentric distances (where the surface temperature is above 190K), the reduction of the sublimation coefficient leads to a significant growth of the surface temperature (up to 20K) and to a change of the temperature distribution near the surface. The heating causes trapped volatiles in the ice to sublimate more actively than simple models predict. The deviation of the sublimation coefficient from unity and the resulting temperature increase can cause crystallization of amorphous water ice at larger heliocentric distances and/or for larger nucleus depth. The increase in temperature associated with the damping of the sublimation coefficient leads to an exponential increase of the corresponding saturation pressure. We note that the covering porous dust layer also induces an increase of the surface temperature of the ice (the so-called "cooking effect"). Thus, one can expect that both phenomena explored in this paper can produce multiple effects leading to an increase of the gas pressure beneath the porous dust layer. If we assume that the gas pressure is the main driver for the breakup of the dust crust, such a pressure increase is a prerequisite for the explanation of dusty outbursts.

\subsection*{Acknowledgements}
Yu. V. Skorov was supported by DFG under grant BI 298/9-1. We thank Sartorius and Millipore for providing us with different filter types.

\bibliographystyle{model2-names}
\bibliography{bib}

\end{document}